# *Ab Initio* Transfer Length Method Simulations of Tunneling Limits in 2D Semiconductors

Tae Hyung Kim, Juho Lee, and Yong-Hoon Kim*

*School of Electrical Engineering, Korea Advanced Institute of Science and Technology (KAIST), 291 Daehak-ro, Yuseong-gu, Daejeon 34141, Korea.*

E-mail: y.h.kim@kaist.ac.kr

**Abstract**

As semiconductor devices approach the sub-2 nm technology node, identifying the quantum-mechanical limits of contact-resistance scaling becomes imperative; however, the transition from thermionic emission to direct tunneling in this deep nanoscale regime remains experimentally inaccessible and theoretically undefined. Herein, we present a systematic first-principles framework to characterize metal/2D-semiconductor interfaces at the atomic scale and identify their intrinsic contact resistance and tunneling limits. Based on large-scale multi-space density functional theory calculations, we perform *ab initio* transmission line model (TLM) analyses for monolayer $MoS_2$ contacted by Sc, Ag, Au, and Pd electrodes in both top-contact and edge-contact geometries. This computational procedure reveals a universal transition in resistance scaling—from metal-induced gap states-mediated direct tunneling in the sub-10 nm regime to thermionic emission at longer channel lengths. The resulting transition length provides a rigorous first-principles measure of the critical tunneling length, establishing a physically grounded metric for assessing contact quality and the source-to-drain tunneling limit of 2D ballistic transistors. Using the *ab initio* TLM method, we further identify optimal contact strategies—top contact with low-work-function metals for n-type operation and edge contact with high-work-function metals for p-type operation. Our study introduces a general computational framework for evaluating and comparing 2D semiconductor contacts and offers practical guidelines for engineering low-resistance, scalable contact technologies for next-generation 2D transistors.

## 1. INTRODUCTION

As conventional silicon transistors approach the limits of geometric scaling, two-dimensional (2D) van der Waals (vdW) semiconductors have emerged as promising channel materials for next-generation field-effect transistors (FETs) [1-4]. In principle, their atomically thin bodies naturally suppress short-channel effects, enabling excellent gate control and allowing sub-100 nm FETs to achieve high on/off ratios and steep subthreshold swings. In practice, however, the practical performance of 2D FETs is critically constrained by electrical contacts, which typically exhibit large contact resistances ($R_c$) that originate from strong Fermi level pinning (FLP) and the resulting Schottky barrier (SB) formation. Consequently, despite notable advances, understanding and optimizing metal/2D-semiconductor interfaces remain central challenges for practical device scaling [1,5-8].

A major difficulty in addressing these challenges has been the persistent mismatch between theoretical predictions and experimental extractions of mobility and contact resistance in 2D semiconductors. First-principles Boltzmann transport studies predict intrinsic phonon-limited mobilities exceeding several hundred cm² V⁻¹ s⁻¹ for monolayer $MoS_2$ and other transition-metal dichalcogenides (TMDCs), yet aggressively scaled experimental devices commonly report mobilities in the 10–50 cm² V⁻¹ s⁻¹ range, particularly for channel lengths below 100 nm [1,9]. Likewise, while simple band alignment arguments suggest that carefully chosen metal work functions should enable low-resistance contacts, experiments consistently observe contact resistances in the kΩ·µm range even in optimized devices [1,5-8].

Conventional equilibrium density functional theory (DFT) calculations are incapable of fully resolving these issues because they describe only static band alignment and do not include non-equilibrium electrostatics or transport physics [10-15]. On the other hand, fully self-consistent DFT–based non-equilibrium Green's function (NEGF) simulations remain computationally prohibitive for large-scale 2D junctions with realistic metal electrodes [16-18]. As a result, the field has lacked a systematic, first-principles-based analysis scheme capable of quantitatively characterizing contact resistance, identifying dominant transport mechanisms, and assessing intrinsic scaling limits in 2D transistors. In particular, finite-bias first-principles calculations that allow transmission line model or transfer length method (TLM) analysis, a standard experimental technique for extracting $R_c$ [1,5,9], have not yet been reported.

In this work, we address this gap by presenting a comprehensive first-principles analysis scheme for metal/2D semiconductor contacts. On the basis of the multi-space constrained-search density functional theory (MS-DFT) formalism[19-23], we perform explicit finite-bias quantum transport calculations for metal–$MoS_2$–metal junctions with top- and edge-contact geometries. By systematically varying the channel length, metal species or work function (Sc, Ag, Au, and Pd), and contact geometry, we extract the contact resistance via *ab initio* TLM analysis (**Fig. 1**a). A key outcome of this procedure is the discovery of a universal transition in resistance scaling, which appears at all the

metal–MoS$_2$ interfaces regardless of the contact geometry. At sub-10-nm channel lengths, the resistance increases exponentially owing to metal induced gap states (MIGS)-mediated direct tunneling, whereas at longer lengths, it increases linearly owing to thermionic emission. The channel length corresponding to this transition provides a rigorous first-principles definition of the effective SB tunneling width, which directly corresponds to the source-to-drain direct-tunneling scaling limit of ballistic 2D FETs (critical tunneling length). By combining these insights, our systematic analysis scheme enables a consistent physical interpretation of the contact resistance, SB formation, and transport mechanisms in 2D devices. Furthermore, it reveals clear design rules for achieving low-resistance contacts: top contact with low-work-function metals for n-type operation and edge contact with high-work-function metals for p-type operation. Beyond MoS$_2$, our framework is broadly applicable to other 2D semiconductors and establishes a foundation for the rational design and benchmarking of contact technologies in extremely scaled 2D transistors.

## 2. METHODS

### 2.1 DFT calculations

All DFT calculations were performed via the SIESTA software package [24] within the local density approximation [25]. While LDA is known to underestimate semiconductor bandgaps, potentially affecting the absolute magnitude of Schottky barrier heights, it reliably captures the trends in electronic coupling, electrostatic potential landscapes, and relative barrier scaling across different metal interfaces. Since the primary focus of this work is to identify the universal transition in resistance scaling and compare the relative efficacy of contact geometries, the LDA formalism provides a computationally efficient and qualitatively robust framework for these comparative analyses. The atomic cores were replaced by Troullier-Martins-type norm-conserving pseudopotentials [26], and double ζ-plus-polarization-level numerical atomic orbital basis sets were employed. For real-space integration, a mesh cutoff of 200 Ry was used. For all the considered junction models, we initially prepared single-electrode interface models via DFT geometry optimizations. In each case, while fixing the geometry of MoS$_2$, we first adjusted the geometry of the metal slab to minimize the mismatch between the lattice constants of MoS$_2$ and the metal (see Fig. S1, Supporting Information). We then scanned the distances between the fixed-geometry metal and MoS$_2$ and identified the minimum energy distance based on the binding energy profile (see Fig. S2, Supporting Information). The interface between the metal and MoS$_2$ was not optimized for comparison between the ideal top- and edge-contact metal-MoS$_2$ structures. Additional computational details are also provided below.

### 2.2 MS-DFT calculations

For finite-bias non-equilibrium electronic structure calculations, we utilized the MS-DFT formalism [19-23] we implemented within the SIESTA code. For the top-contacted metal-MoS$_2$-metal junction models (Fig. 1a, top panel; see also Fig. S3, Supporting Information), we used supercells with dimensions of $x$ = 5.4 Å × $y$ = 35 Å × $z$ = $Z$ Å with the periodic boundary condition (PBC) along the $x$ direction and increasing $Z$ proportionally with $L_{\text{ch}}$. We employed Monkhorst-Pack $\bm{k}$-points with dimensions of 12×1×1. For the edge-contact metal-MoS$_2$-metal junction models, we used supercells with dimensions of 5.4 Å × $Y$ Å × $Z$ Å, where $Y$ ranges between 18 Å and 19 Å depending on the metal species, and $Z$ increases with $L_{\text{ch}}$. For edge-contacted junction models (Fig. 1a, bottom panel; see also Fig. S3, Supporting Information), we additionally adopted the PBC along the $y$ direction and sampled Monkhorst-Pack $\bm{k}$-point grids of 12×3×1. We examined several edge terminations of MoS$_2$ and found that the armchair MoS$_2$ edge and S-terminated zigzag MoS$_2$ edge exhibit similar junction resistance and resistance transition features, whereas the Mo-terminated zigzag MoS$_2$ edge present the highest junction resistance values in the long-$L_{\text{ch}}$ resistance scaling regime (Fig. S4, Supporting Information). Accordingly, we selected the armchair edge termination geometry for the final data included in the manuscript.

After obtaining the non-equilibrium device electronic structures within the MS-DFT formalism, we calculated the transmissions as post-processing steps based on the matrix Green's function formalism [27,28]. The left/right-electrode surface Green's functions $\bm{g}_{\text{s,L/R}}$ were obtained from separate DFT calculations for periodic top-contact metal-TMDC structures for top-contact junction models and periodic bulk metal structures for edge-contact junction models. The effects of coupling to L/R electrodes are introduced through self-energies $\bm{\Sigma}_{\text{L/R}} = \bm{\tau}_{\text{L/R}}\,\bm{g}_{\text{s,L/R}}\,\bm{\tau}_{\text{L/R}}^{+}$, where $\bm{\tau}_{\text{L/R}}$ is the electrode L/R-channel coupling matrix. Finally, the transmission spectra were calculated according to [29-31]

$$T(E;V_{\text{b}}) = \text{Tr}[\bm{\Gamma}_{\text{L}}(E;V_{\text{b}})\bm{G}(E;V_{\text{b}})\bm{\Gamma}_{\text{R}}(E;V_{\text{b}})\bm{G}^{+}(E;V_{\text{b}})], \quad (1)$$

where $\bm{\Gamma}_{\text{L/R}} = i(\bm{\Sigma}_{\text{L/R}} - \bm{\Sigma}_{\text{L/R}}^{+})$ are the broadening matrices and where $\bm{G}$ is the retarded Green's function for the electrode-coupled channel.

## 3. RESULTS AND DISCUSSION

### 3.1 Formation of Schottky barriers at metal-MoS$_2$ contacts

For monolayer MoS$_2$ in contact with Sc, Ag, Au, and Pd metal electrodes with varying work functions ($\phi_{\text{Sc}} = 3.43$ eV, $\phi_{\text{Ag}} = 4.40$ eV, $\phi_{\text{Au}} = 5.11$ eV, and $\phi_{\text{Pd}} = 5.69$ eV) in the top and edge contact geometries, we prepared metal-MoS$_2$-metal junction models through elaborate procedures (see **Fig. S1**--**S4**, Supporting Information) and carried out large-scale DFT and MS-DFT calculations by varying the MoS$_2$ channel length $L_{\text{ch}}$. As an example, we compare the calculated current density–bias voltage ($J$–$V_{\text{b}}$) characteristics of $L_{\text{ch}} = \sim 9$ nm MoS$_2$ junctions prepared by top- and edge-contacting low-work-function Ag (Fig. 1b, left panel) and high-work-function Pd metal electrodes (Fig. 1b, right panel). The $J$ values under the bias voltage $V_{\text{b}} = (\mu_L - \mu_R)/e$, where $\mu_{\text{L/R}}$ is the chemical potential of the left/right (L/R) electrode, were calculated via the Landauer–Büttiker formula [29-31],

$$J(V_b) = \frac{2e}{hW} \int_{-\infty}^{+\infty} T(E; V_b)(f_L - f_R)dE, \quad (2)$$

where $W$ is the channel width, $T(E; V_b)$ is the transmission spectrum at finite $V_b$, and $f_{L/R}(E) = 1/\{1 + \exp[(E - \mu_{L/R})/k_B T]\}$ is the Fermi-Dirac distribution function for the L/R electrode. Throughout this work, we fixed the temperature at $T = 300$ K. All ideal metal-TDMC junctions form rectifying contacts or Schottky barrier diodes, and the Fowler-Nordheim (FN) or $\ln(J/V_b^2)$ vs. $1/V_b$ plots presented in Fig. 1c show that FN tunneling becomes the dominant transport mechanism at approximately $V_b = 0.5$ V. We note that larger currents are obtained with the top contact configuration for Ag but with the edge contact counterpart for Pd. Overall, the current densities of the four MoS$_2$ junctions are in the order of Ag-top > Pd-edge > Ag-edge > Pd-top metal species-contact geometry combinations.

To understand the complex dependence of the long-$L_{ch}$ regime metal-MoS$_2$ contact resistance on the combination of the metal species and contact geometry, taking the Ag top/edge contacts as a representative case, we discuss the equilibrium interfacial electronic structures. In the left panels of **Fig. 2**a, b, together with the junction atomic models (top), we first show the local density of states (LDOS) of the top-contact and edge-contact Ag-MoS$_2$ junctions, respectively (bottom). Once the band alignments are established at the Ag-MoS$_2$ interfaces, minimal band bending occurs at the considered length scales (~ 10 nm) along the lateral $z$-direction for both the top and edge contact configurations. Additionally, we note that while $n$-type Schottky contacts are formed with a top-contact configuration, weak $p$-type character is achieved with an edge-contact geometry [16,17]. The utilization of metal-MoS$_2$-metal junction models makes the estimation of SBs more straightforward, particularly for the top-contact configuration [15]. The Mo 4$d$ projected bands of MoS$_2$ approximately 1 nm away from the Ag-MoS$_2$ interfaces (upward triangles in the atomic models) are shown together with the LDOS plots, from which we obtained the electron SB height (SBH) $\phi_{Bn} = +0.32$ eV and the hole SBH $\phi_{Bp} = -0.69$ eV for the top and edge Ag-MoS$_2$ contacts, respectively.

Evaluating the $\phi_{Bn}$ values for other metal electrode cases (see **Note S1**, **Fig. S5--S7**, and **Table S1**, Supporting Information), we consistently observed $n$-type and $p$-type band alignments for the top-contact and edge-contact geometries, respectively. We then find that as the metal work function increases, $\phi_{Bn}$ increases for the $n$-type top contact case, whereas $\phi_{Bp}$ decreases for the $p$-type edge contact counterpart. The trends can be quantified in terms of the pinning or slope parameter $S = d\phi_{Bn}/d\phi_{metal}$ ($S = 0$ and $S = 1$ correspond to the fully pinned Bardeen limit and fully unpinned Mott–Schottky limit, respectively), and we obtain $S = 0.32$ and $S = 0.21$ for the top-contact and edge-contact geometries, respectively (Fig. S7, Supporting Information). Importantly, the ordering of the resistances of the Ag-Pd-top-edge-contacted monolayer MoS$_2$ junctions shown in Fig. 1 (Ag-top < Pd-edge < Ag-edge < Pd-top) follows that of their SBH magnitudes ($\phi_{Bn} = 0.32$ eV for Ag-top, $|\phi_{Bp}| = 0.56$ eV for Pd-edge, $|\phi_{Bp}| = 0.69$ eV for Ag-edge, and $\phi_{Bn} = 0.88$ eV for Pd-top).

The band alignment and SBH formation can be understood in terms of charge density differences or interface dipoles formed at Ag-MoS$_2$ contacts [15] $\delta\rho = \rho_{Ag-MoS_2} - (\rho_{Ag} + \rho_{MoS_2})$ and corresponding $\delta\rho$-induced plane-averaged Hartree electrostatic potential differences $\delta\bar{V}_H = \bar{V}_H^{Ag-MoS_2} - (\bar{V}_H^{Ag} + \bar{V}_H^{MoS_2})$ (Fig. 2, middle panels). In both the top-contact and edge-contact cases, $\delta\rho$ forms within a range of a few angstroms, leading to abrupt shifts in $\delta\bar{V}_H$ at the interfaces. The increase in $\delta\rho$ or $\delta\bar{V}_H$ determines the interfacial band alignment, which is characterized by the Fermi level pinning effect [5-7] or the pinning of the Fermi level ($E_F$) of Ag to the charge neutrality level (CNL) of MoS$_2$. Owing to the anisotropic crystal structure, the electron affinity level ($\chi$) and CNL ($\phi^*$) values of the top and edge sides of van der Waals 2D crystals are very different [32]. Once the top- and edge-side $\chi$ values of monolayer MoS$_2$ ($\chi_{top} = 4.28$ eV and $\chi_{edge} = 3.93$ eV) are obtained from the defining equation for $\phi_{Bn}$ in terms of the $S$, $\phi^*$, and $\chi$ values,

$$\phi_{Bn} = S(\phi_{metal} - \phi^*_{top/edge}) + (\phi^*_{top/edge} - \chi_{top/edge}), \quad (3)$$

we obtained the CNL of the MoS$_2$ top side ($\phi^*_{edge} = 5.37$ eV) significantly lower than that of the MoS$_2$ top side ($\phi^*_{top} = 4.69$ eV; Note S1 and Fig. S5, Supporting Information) [16,17]. Nevertheless, both CNLs lie below the Ag work function ($\phi_{Ag} = 4.40$ eV). Consequently, electrons transfer from Ag to MoS$_2$ in both the top and edge contact configurations (with different strengths), resulting in electron accumulation at MoS$_2$ (red regions in $\delta\rho$) and positive offsets of $\delta\bar{V}_H$ at the Ag-MoS$_2$ top (+0.29 eV) and edge (+0.72 eV) contacts (Fig. 2 middle and right panels). These different $\delta\bar{V}_H$ offsets explain the $n$-type and weak $p$-type SBH formations for the Ag-MoS$_2$ top and edge contact configurations, respectively, as observed in the LDOS (Fig. S6, Supporting Information).

### 3.2 First-principles TLM analysis and the transition in resistance scaling

To systematically analyze the dependence of $R_c$ on the metal species and contact geometry, we performed first-principles TLM calculations and extracted resistance $R = (J/V_b)^{-1}$ values with increasing $L_{ch}$ under zero-bias equilibrium and $V_b = 0.1$ V non-equilibrium conditions. In **Fig. 3**a,b, we present the finite-bias TLM data for the top-contact (left panels; blue filled squares) and edge-contact (right panels; red filled right triangles) Ag-MoS$_2$-Ag and Pd-MoS$_2$-Pd junctions, respectively, both on the logarithmic scale (top panels) and on the linear scale (bottom panels). As the most notable common feature, we observe sharp transitions in the resistance scaling curves. Also observed with carbon nanotubes and thus must be a common feature for SB interfaces [33,34], the transition in the $L_{ch}$ scaling of resistance arises because the dominant transport mechanism changes from MIGS-mediated direct tunneling (DT) to thermionic emission (TE). The identified contact resistance transition channel length ($L_{ch}^T$) corresponds to the effective SB tunneling width and, because the prominent source-to-drain DT will disable gate control or make FETs ineffective, can be considered the fundamental scaling limit of TMDC FETs.

To understand the resistance scaling transition behavior, we first discuss the short-$L_{ch}$ resistance scaling regime, where the resistance increases exponentially with increasing $L_{ch}$ (solid trend lines in the top panels of Fig. 3a,b). The exponential increase in resistance arises because the dominant charge transport mechanism is the MIGS-mediated DT. Fitting the exponential growth trend with [33,34]

$$R(L_{ch})W = 2R_c^{short}W \exp(kL_{ch}), \quad (4)$$

where $k$ is a growth constant. We extracted the contact resistance in the short-$L_{ch}$ regime ($R_c^{short}W$) from the $y$-axis intercept of the log-scale $RW$-$L_{ch}$ curve. We find that the calculated $R_c^{short}W$ values from edge-contacted junctions are consistently lower than those from their top-contacted counterparts by approximately two orders of magnitude (see **Fig. S8**, Supporting Information). However, given the dominance of DT or limited gate controllability in this short-$L_{ch}$ resistance scaling regime, the nature of contacts providing low $R_c^{short}$ will be irrelevant for the practical realization of FETs.

On the other hand, in the long-$L_{ch}$ resistance scaling regime, the resistance values no longer follow exponential growth behavior and increase linearly as $L_{ch}$ increases (solid trend lines in the bottom panels of Fig. 3a,b). This implies that the dominant low-bias transport mechanism changes from DT to TE. Accordingly, the width-normalized resistance can be expressed in terms of the conventional TLM equation for 2D semiconductors [5-7]:

$$R(L_{ch})W = 2R_c^{long}W + \rho_{2D}L_{ch}, \quad (5)$$

where $\rho_{2D}$ is the resistivity of the 2D channel. By extracting the width-normalized $R_c^{long}W$ values from the $y$-axis intercepts of the normal-scale $RW$-$L_{ch}$ curves, as indicated earlier in Fig. 1b, we observe that the $R_c^{long}W$ values depend on both the metal species and contact geometry. Quantitatively, they are in the order of Ag-top (1.2×10$^5$ kΩ·µm) < Pd-edge (1.8×10$^8$ kΩ·µm) < Ag-edge (4.4×10$^{11}$ kΩ·µm) < Pd-top (1.1×10$^{14}$ kΩ µm) combinations, matching the order of SBHs. We also find that, whereas the ordering of the $L_{ch}^T$ of SB width values of the Ag-top (3.5 nm) < Pd-edge (5.6 nm) < Ag-edge (6.9 nm) ≈ Pd-top (6.7 nm) combinations closely matches those of the $R_c^{long}W$ and SBHs, they are not exactly equivalent.

We repeated TLM analyses based on $V_b = 0.1$ V MS-DFT calculations for all combinations of metal species (Sc, Ag, Au, and Pd) and contact geometries (top and edge). As summarized in Fig. 3c, we then observe the transition in resistance scaling for all cases and the strong correlations between $R_c^{long}W$, SBH, and $L_{ch}^T$ variations.

As discussed earlier, top-contacted (edge-contacted) MoS$_2$ junctions form $n$-type ($p$-type) band alignments, and accordingly, the SBH increases (decreases) with the metal work function (Fig. 3c left panel) and produces increasing (decreasing) $R_c^{long}W$-metal work function curves (Fig. 3c middle panel).

Overall, we found that the Sc top-contacted (Pd edge-contacted) junction produced the lowest $n$-type ($p$-type) $R_c^{long}W$ value of 2.3×10$^2$ kΩ·µm (1.8×10$^8$ kΩ·µm). While the correlation between $R_c^{long}W$ and SBH is strong than that between $R_c^{long}W$ and $L_{ch}^T$ (Fig. 3c, right panel), $L_{ch}^T$ or the critical tunneling length (the shortest $n$-type $L_{ch}^T$ =3.1 nm with the Sc top contact and the shortest $p$-type $L_{ch}^T$ =5.6 nm with the Pd edge contact) provides additional crucial information on the scaling limits of MoS$_2$ FETs enforced by direct tunneling.

In the long-$L_{ch}$ resistance scaling regime, we additionally evaluated the electron mobility $\mu_e$ and hole mobility $\mu_h$ values according to

$$\rho_{2D} = (en_e\mu_e + en_h\mu_h)^{-1}, \quad (6)$$

where $n_e$ and $n_h$ are the electron and hole densities, respectively (see **Note S2**, **Fig. S9**, and **Table S2**, Supporting Information). The obtained $\mu_e$ values from the top-contact junctions range between 1.9 and 66.4 cm$^2$V$^{-1}$s$^{-1}$, and the $\mu_h$ values from the edge-contact junctions range between 4.2 and 74.5 cm$^2$V$^{-1}$s$^{-1}$. We note that, e.g., our calculated nobility value of 66.4 cm$^2$V$^{-1}$s$^{-1}$ obtained for the Au-top contact case is in good agreement with the experimental value of 59 cm$^2$V$^{-1}$s$^{-1}$ extracted from the TLM measurement of Au top-contacted monolayer MoS$_2$ in the low carrier density regime [9]. Importantly, we note the necessity of explicit finite-bias calculations to produce correct long-$L_{ch}$ resistance scaling behaviors. The $V_b$ = 0.1 V TLM data approximately evaluated with the zero-bias transmissions $T(E; V_b = 0$ V$)$ are shown in the bottom panels of Fig. 3a,b (blue empty squares and red empty right triangles together with dashed trend lines). We then find consistently underestimated $R_c^{long}W$ and $\rho_{2D}$ values, demonstrating how equilibrium approximations that fundamentally misinterpret the electrostatic landscape of ultra-short channels lead to the persistent mobility mismatch observed in literature.

### 3.3 Mechanisms of the transition in resistance scaling

We explained that the transition in the $L_{ch}$ scaling of resistance originates from the change in the dominant transport mechanism from MIGS-mediated DT (exponentially increasing resistance) to TE (linearly increasing resistance). Finally, we discuss the mechanisms of the resistance scaling transition in more detail by decomposing the $V_b = 0.1$ V current from the metal-MoS$_2$-metal junctions into DT, FN tunneling, and TE components according to [35,36]

$$J^{DT}(V_b) = \frac{2e}{hW} \int_{S-VBM}^{D-CBM} T(E; V_b)(f_L - f_R)dE,$$

$$J^{FN}(V_b) = \frac{2e}{hW} \left[ \int_{D-CBM}^{S-CBM} T(E; V_b)(f_L - f_R)dE \right.$$
$$\left. + \int_{D-VBM}^{S-VBM} T(E; V_b)(f_L - f_R)dE \right], \quad (7)$$

$$J^{\text{TE}}(V_{\text{b}}) = \frac{2e}{hW}\left[\int_{\text{S-CBM}}^{+\infty} T(E;V_{\text{b}})(f_{\text{L}}-f_{\text{R}})dE \right.$$
$$\left. + \int_{-\infty}^{\text{D-VBM}} T(E;V_{\text{b}})(f_{\text{L}}-f_{\text{R}})dE \right].$$

Here, l-CBM (r-CBM) and l-VBM (r-VBM) indicate the CBM and VBM energy levels at the L (R) position, respectively. Reflecting the negligible band bending along the MoS$_2$ channel, we assigned the L-CBM (R-CBM) and L-VBM (R-VBM) levels by shifting the equilibrium CBM and VBM energy levels as identified above (Fig. 2) by $+\mu_{\text{S}}/2$ ($-\mu_{\text{D}}/2$), respectively (indicated as dotted lines in **Fig. 4,** first left panels). Accordingly, the resistance values can be decomposed into DT resistance $R^{\text{DT}} = (J^{\text{DT}}/V_{\text{b}})^{-1}$, FN tunneling resistance $R^{\text{FN}} = (J^{\text{FN}}/V_{\text{b}})^{-1}$, and TE resistance $R^{\text{TE}} = (J^{\text{TE}}/V_{\text{b}})^{-1}$ components, with a total resistance of $R = (R^{\text{DT}^{-1}} + R^{\text{FN}^{-1}} + R^{\text{TE}^{-1}})^{-1}$.

In Fig. 4, we also present the LDOS (first left panels), transmission $T(E;V_{\text{b}} = 0.1\text{ V})$ (second left panels), differences in L/R-electrode Fermi-Dirac distributions $(f_{\text{L}}-f_{\text{R}})(E;V_{\text{b}} = 0.1\text{ V})$ (third left panels), and energy-resolved currents $T \times (f_L-f_R)(E;V_{\text{b}} = 0.1\text{ V})$ (right panels) obtained from short-$L_{\text{ch}}$ (2.4 nm) and long-$L_{\text{ch}}$ (10.2 nm) top-contact Ag-MoS$_2$-Ag junctions with $L_{\text{ch}}^{\text{T}} = 3.5$ nm and short-$L_{\text{ch}}$ (3.1 nm) and long-$L_{\text{ch}}$ (10.3 nm) edge-contacted Pd-MoS$_2$-Pd junctions with $L_{\text{ch}}^{\text{T}} = 5.6$ nm. Then, in the short-$L_{\text{ch}}$ cases, we confirmed that the dominant transport mechanism is the DT mediated by MIGS (the purple regions in the $T \times (f_{\text{L}}-f_{\text{R}})$ panels of Fig. 4a,c).

However, as $L_{\text{ch}}$ increases above $L_{\text{ch}}^{\text{T}}$ (Fig. 4b,d), the crosstalk between the L- and R-contact MIGS across the MoS$_2$ channel becomes insignificant, leading to negligible DT contributions and the exponential growth of the resistance with increasing $L_{\text{ch}}$ in the $L_{\text{ch}} < L_{\text{ch}}^{\text{T}}$ regime (Fig. 4e,f). In the $L_{\text{ch}} > L_{\text{ch}}^{\text{T}}$ regime, FN tunneling and TE outside the bias window dominate carrier transport (orange and magenta regions, respectively, in the panels of Fig. 4b,d $T \times (f_{\text{L}} - f_R)$**)**. The TE component accounts for carrier transport in both the top and edge contact cases (magenta regions in the right panels of Fig. 4b,d), explaining the linear increase in resistance in accordance with Equation (4) (Fig. 4e,f). Here, we also note that while the TE contribution is dominant in the top-contacted Ag-MoS$_2$-Ag junction, the FN tunneling contribution is comparable to the TE component in the edge-contacted Pd-MoS$_2$-Pd counterpart. Because FN tunneling could hinder electrostatic gating control like DT, this might represent an inherent limitation of adopting an edge contact geometry in view of FET scaling.

## 4. CONCLUSION

In summary, for sub-15 nm $L_{\text{ch}}$ monolayer MoS$_2$ junction models established from top- and edge-contacting Sc, Ag, Au, and Pd electrodes, we performed extensive MS-DFT calculations to conduct, for the first time, *ab initio* TLM analyses of 2D semiconductors on the basis of finite-bias quantum transport calculations. A key outcome of the procedure was the universal transition feature in the $L_{\text{ch}}$ scaling of resistance, which occurs between approximately 3 nm and 9 nm depending on the metal species–contact geometry combination. Originating from a shift in the dominant carrier transport mechanism from direct tunneling in the short-$L_{\text{ch}}$ regime to thermionic emission in the long-$L_{\text{ch}}$ regime, the resistance scaling transition length $L_{\text{ch}}$ corresponds to the critical tunneling length. We thus established a rigorous scheme to measure the effective SB tunneling width or the direct tunneling-determined scaling limits of ballistic 2D FETs.

Our results provide specific design guidelines: low-$R_c$ *n*-type contacts are best achieved via top-contacts with low-work function metals (e.g. Sc, Ag), while *p*-type contacts are best achieved via edge-contacts with high-work function metals (e.g. Au, Pd). These findings also point toward a novel asymmetric design direction for monolithic 2D CMOS logic, where optimal complementary performance is achieved by pairing top contacts for electron injection with edge contacts for hole injection. Such a hybrid scheme effectively circumvents the trade-offs inherent in single-geometry contact technologies, offering a pathway to balanced high-performance 2D logic gates. Filling a critical gap in experimental capabilities and equilibrium DFT studies in probing sub-10 nm transport regimes, we expect our *ab initio* TLM approach will play a valuable role for the rational design of contacts in future ultrascaled low-dimensional electronics [37,38].

## Acknowledgements

This research was supported by the National Research Foundation of Korea (Grant Nos. 2022K1A3A1A91094293, 2023R1A2C2003816, and RS-2023-00253716) and Samsung Electronics. This study was also supported by the Korea Advanced Institute of Science and Technology grant BK21 Plus. Computational resources were provided by the KISTI Supercomputing Center (KSC-2022-CRE-0258).

## Author Contributions

T.H.K. performed the calculations, analyzed the results, and prepared a draft of the manuscript. J.L. assisted MS-DFT calculations and data analysis. Y.-H.K. formulated and oversaw the project, analyzed the results, and wrote the manuscript with input from T.H.K.

## REFERENCES

[1] S. Das *et al*., Nat. Electron. **4**, 786 (2021).
https://doi.org/10.1038/s41928-021-00670-1
[2] T. Wei *et al*., iScience **25**, 105160 (2022).
https://doi.org/10.1016/j.isci.2022.105160
[3] K. S. Kim *et al*., Nat. Nanotechnol. **19**, 895 (2024).
https://doi.org/10.1038/s41565-024-01695-1
[4] H. Li *et al*., Adv. Funct. Mater. **34**, 2402474 (2024).
https://doi.org/10.1002/adfm.202402474
[5] S. B. Mitta *et al*., 2D Mater. **8**, 012002 (2020).
https://doi.org/10.1088/2053-1583/abc187
[6] X. Liu, M. S. Choi, E. Hwang, W. J. Yoo, and J. Sun, Adv. Mater. **34**, 2108425 (2022). https://doi.org/10.1002/adma.202108425
[7] Y. Wang and M. Chhowalla, Nat. Rev. Phys. **4**, 101 (2022).
https://doi.org/10.1038/s42254-021-00389-0
[8] S. Zeng, C. Liu, and P. Zhou, Nat. Rev. Electr. Eng. **1**, 335 (2024).
https://doi.org/10.1038/s44287-024-00045-6

[9] Z. Cheng *et al.*, Nat. Electron. **5**, 416 (2022). https://doi.org/10.1038/s41928-022-00798-8
[10] I. Popov, G. Seifert, and D. Tomanek, Phys. Rev. Lett. **108**, 156802 (2012). https://doi.org/10.1103/PhysRevLett.108.156802
[11] J. Kang, W. Liu, D. Sarkar, D. Jena, and K. Banerjee, Phys. Rev. X **4**, 031005 (2014). https://doi.org/10.1103/PhysRevX.4.031005
[12] C. Gong, L. Colombo, R. M. Wallace, and K. Cho, Nano Lett. **14**, 1714 (2014). https://doi.org/10.1021/nl403465v
[13] Y. Liu, P. Stradins, and S. H. Wei, Sci. Adv. **2**, e1600069 (2016). https://doi.org/10.1126/sciadv.1600069
[14] J. Shim *et al.*, Adv. Mater. **28**, 5293 (2016). https://doi.org/10.1002/adma.201506004
[15] B.-K. Kim *et al.*, npj 2D Mater. Appl. **5**, 9 (2021). https://doi.org/10.1038/s41699-020-00191-z
[16] D. Çakır and F. M. Peeters, Phys. Rev. B **89**, 245403 (2014). https://doi.org/10.1103/PhysRevB.89.245403
[17] K. Parto *et al.*, Phys. Rev. Appl. **15**, 064068 (2021). https://doi.org/10.1103/PhysRevApplied.15.064068
[18] A. Afzalian, npj 2D Mater. Appl. **5**, 1 (2021). https://doi.org/10.1038/s41699-020-00181-1
[19] J. Lee, H. S. Kim, and Y.-H. Kim, Adv. Sci. **7**, 2001038 (2020). https://doi.org/10.1002/advs.202001038
[20] J. Lee, H. Yeo, and Y.-H. Kim, Proc. Natl. Acad. Sci. U.S.A. **117**, 10142 (2020). https://doi.org/10.1073/pnas.1921273117
[21] T. H. Kim, J. Lee, R.-G. Lee, and Y.-H. Kim, npj Comput. Mater. **8**, 50 (2022). https://doi.org/10.1038/s41524-022-00731-9
[22] J. Lee, H. Yeo, R.-G. Lee, and Y.-H. Kim, npj Comput. Mater. **10**, 1 (2024). https://doi.org/10.1038/s41524-024-01242-5
[23] Y.-H. Kim and R.-G. Lee, IEEE Nanotechnol. Mag. **18**, 4 (2024). https://doi.org/10.1109/mnano.2024.3475888
[24] J. M. Soler *et al.*, J. Phys. Condens. Matter **14**, 2745 (2002). https://doi.org/10.1088/0953-8984/14/11/302
[25] D. M. Ceperley and B. J. Alder, Phys. Rev. Lett. **45**, 566 (1980). https://doi.org/10.1103/PhysRevLett.45.566
[26] N. Troullier and J. L. Martins, Phys. Rev. B **43**, 1993 (1991). https://doi.org/10.1103/PhysRevB.43.1993
[27] Y.-H. Kim, S. S. Jang, Y. H. Jang, and W. A. Goddard III, Phys. Rev. Lett. **94**, 156801 (2005). https://doi.org/10.1103/PhysRevLett.94.156801
[28] Y.-H. Kim, J. Tahir-Kheli, P. A. Schultz, and W. A. Goddard III, Phys. Rev. B **73**, 235419 (2006). https://doi.org/10.1103/PhysRevB.73.235419
[29] S. Datta, *Quantum Transport: Atom to Transistor* (Cambridge University Press, Cambridge, 2005).
[30] M. P. Anantram, M. S. Lundstrom, and D. E. Nikonov, Proc. IEEE **96**, 1511 (2008). https://doi.org/10.1109/JPROC.2008.927355
[31] D. Vasileska, S. M. Goodnick, and G. Klimeck, *Computational Electronics* (Springer, 2017).
[32] Y. Guo, D. Liu, and J. Robertson, ACS Appl. Mater. Interfaces **7**, 25709 (2015). https://doi.org/10.1021/acsami.5b06897
[33] H. S. Kim, H. S. Kim, G. I. Lee, J. K. Kang, and Y.-H. Kim, MRS Commun. **2**, 91 (2012). https://doi.org/10.1557/mrc.2012.14
[34] Y.-H. Kim and H. S. Kim, Appl. Phys. Lett. **100**, 213113 (2012). https://doi.org/10.1063/1.4721487
[35] S. Das, A. Prakash, R. Salazar, and J. Appenzeller, ACS Nano **8**, 1681 (2014). https://doi.org/10.1021/nn406603h
[36] F. Ahmed, M. S. Choi, X. Liu, and W. J. Yoo, Nanoscale **7**, 9222 (2015). https://doi.org/10.1039/C5NR01044F
[37] F. Léonard and A. A. Talin, Nat. Nanotechnol. **6**, 773 (2011). https://doi.org/10.1038/nnano.2011.196
[38] H. Yang *et al.*, Nature **606**, 663 (2022). https://doi.org/10.1038/s41586-022-04768-0

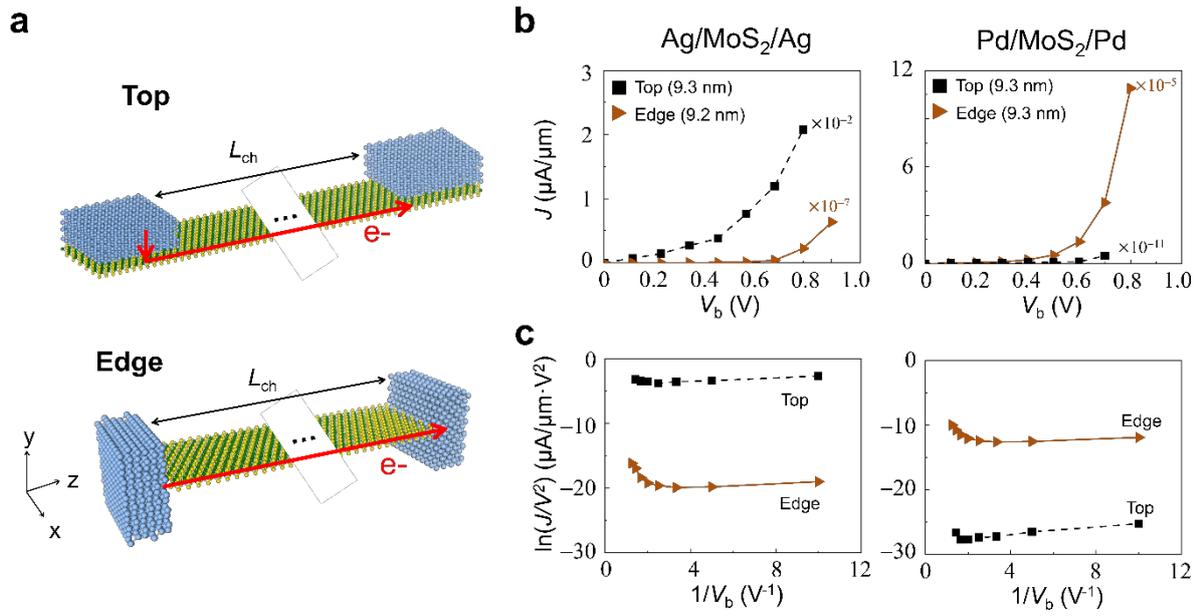

**Figure 1.** Contact geometries and their JV characteristics. (a) Schematics of the top-contact (top panel) and edge-contact (bottom panel) metal-MoS$_2$-metal junction models. (b) Current density–bias voltage ($J - V_b$) curves (top panels) and (c) $\ln(J/V_b^2)$–$1/V_b$ FN curves obtained from Ag-$L_{ch} \approx 8$ nm MoS$_2$-Ag (left panels) and Pd-$L_{ch} \approx 9$ nm MoS$_2$-Pd (right panels) junction models. The black squares and brown triangles on the right represent the data from the top-contact and edge-contact junction models, respectively.

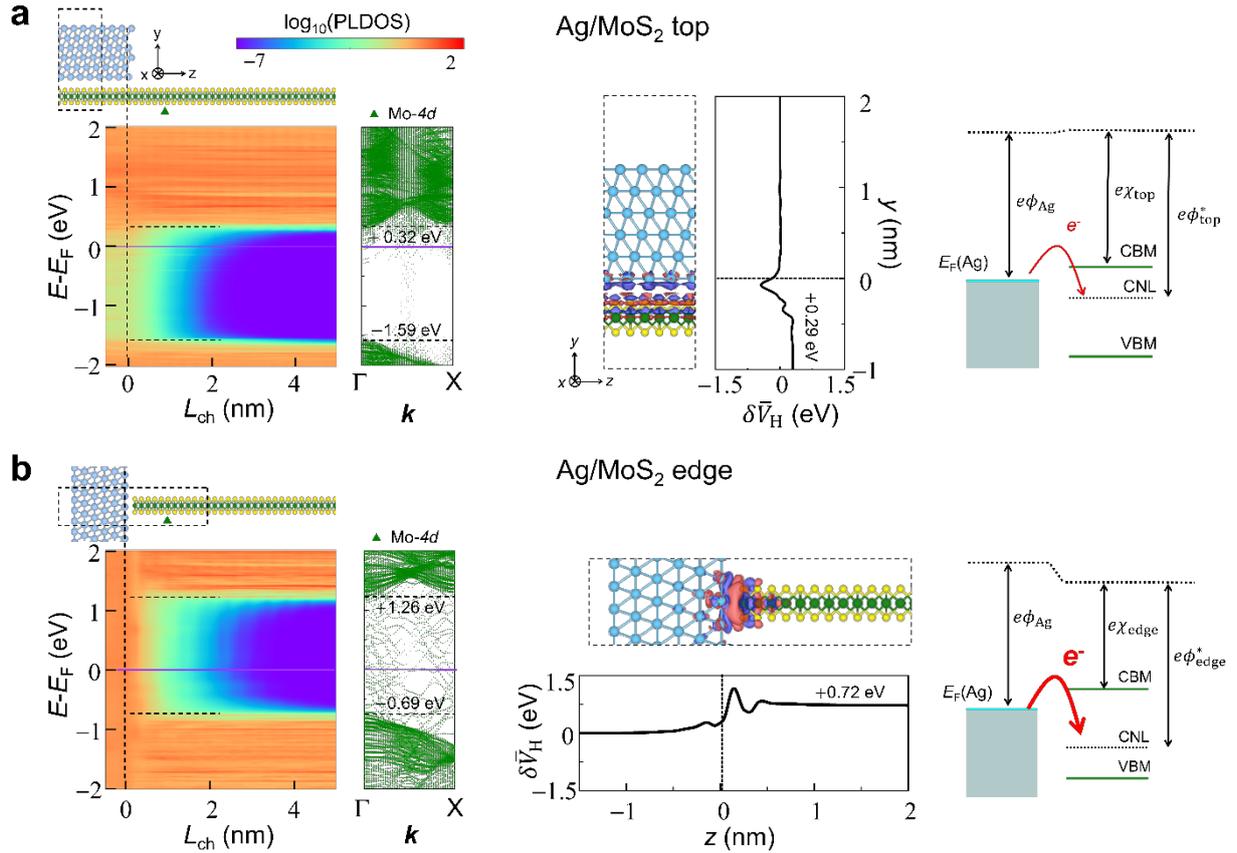

**Figure 2.** Equilibrium electronic structures of the interfacial regions of Ag-MoS$_2$-Ag junctions. For the (a) top and (b) edge Ag-MoS$_2$-Ag contacts, in the left panels, we show the LDOS (bottom-left panels) and the projected bands of the Mo-*4d* orbitals approximately 1 nm away from the Ag-MoS$_2$ interfaces (bottom-right panels). The upper and lower black dashed lines in the LDOS figures indicate the CBM and VBM of MoS$_2$, respectively, and the purple solid lines represent the Fermi level of Ag. In the middle panels, we show $\delta\rho$ overlaid over Ag-MoS$_2$ contact atomic structures and $\delta\bar{V}_H$. The atomic structures are zoomed-in views of the regions indicated by black dashed boxes in the left-top panels. The red and blue areas indicate the charge accumulation and depletion regions, where the iso-surface level of $\delta\rho$ is $6.75 \times 10^{-3}$ e · Å$^{-3}$. In the right panels, we show schematics of the Ag and MoS$_2$ band levels before junction formation. The red arrows indicate the net charge transfer directions when contacts are formed, and the arrow thicknesses are schematically proportional to the amount of net charge transfer.

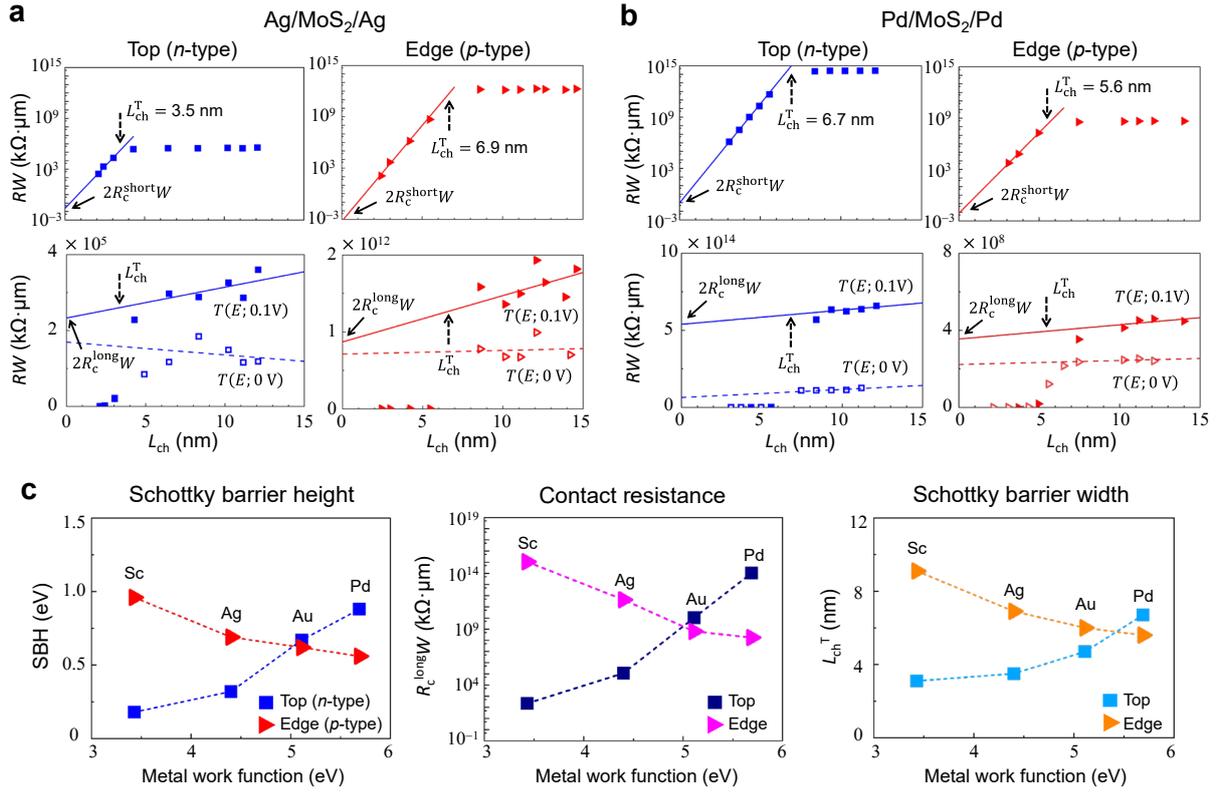

**Figure 3.** First-principles TLM identification of the critical tunneling length. (a), (b) Resistances of top-contact Ag-MoS$_2$-Ag (left panels in (a)), edge-contact Ag-MoS$_2$-Ag (right panels in (a)), top-contact Pd-MoS$_2$-Pd (left panels in (b)), and edge-contact Pd-MoS$_2$-Pd (right panels in (b)) configurations as a function of $L_{ch}$ are shown on the logarithmic (upper panels) and linear (lower panels) scales. The blue filled squares (red filled right triangles) represent $V_b$ = 0.1 V TLM data from top-contact (edge-contact) junctions. In the upper panels, the solid lines indicate the trendlines of the three data points in the short-$L_{ch}$ scaling regime. The same $V_b$ = 0.1 V TLM data are reproduced in the lower panels, together with the solid trendlines for the long-$L_{ch}$ scaling regime (for more details, refer to Figure S8, Supporting Information). In the lower panels, the corresponding data obtained from equilibrium $T(E; 0\text{ V})$ curves (blue empty squares and red empty right triangles) are presented together with their trendlines (dashed lines). In (a) and (b), the short-$L_{ch}$ scaling-to-long-$L_{ch}$ scaling transition channel lengths ($L_{ch}^T$) are indicated by up arrows. (c) Variations in the SBH, $R_c^{TE}$, and $L_{ch}^T$ of MoS$_2$ junctions depending on the contact geometry and metal work function. The filled squares (right triangles) represent the data from top-contact $p$-type (edge-contacted $n$-type) junctions.

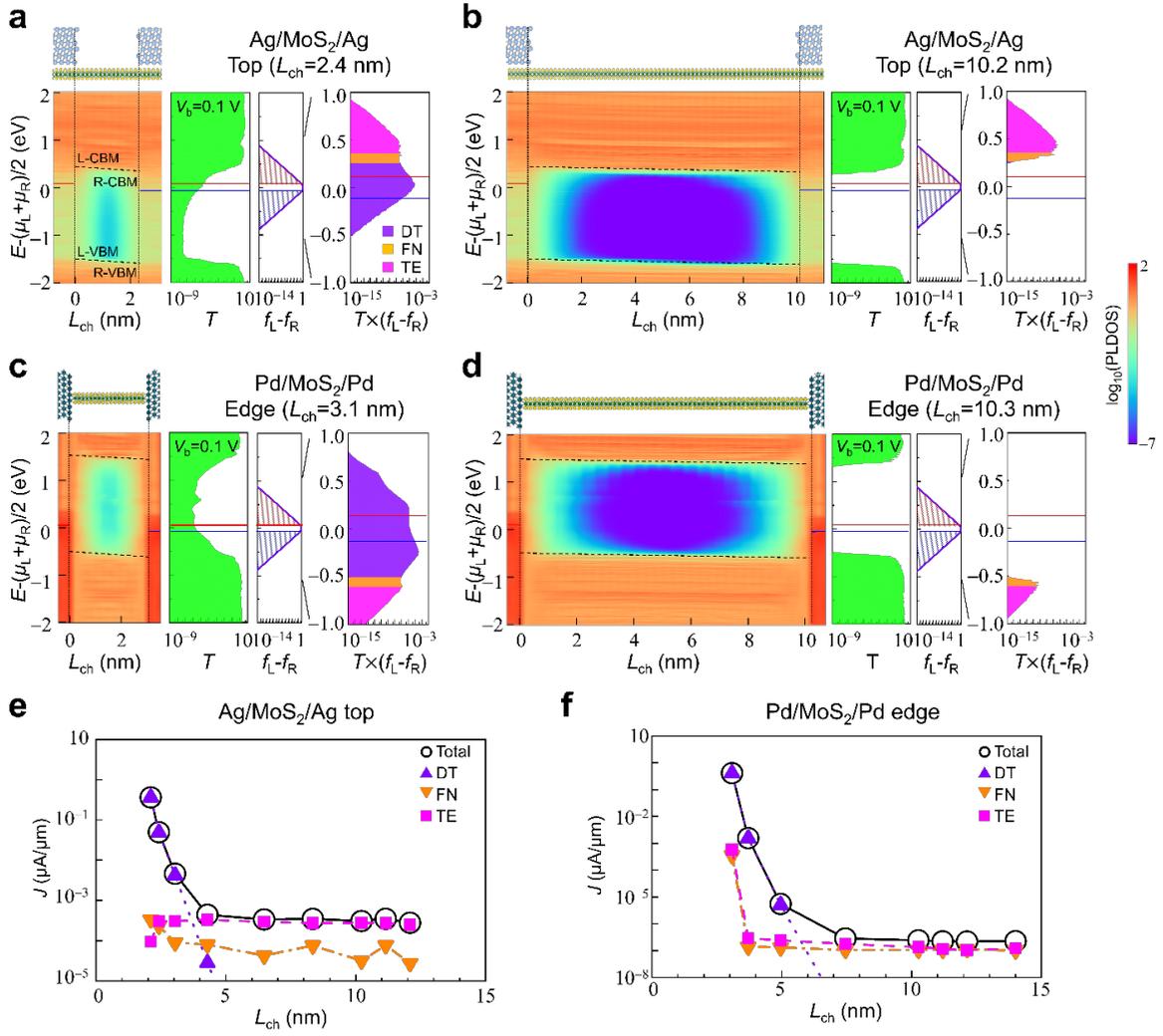

**Figure 4.** Electronic structures and dominant transport mechanisms of short and long MoS$_2$ channels. At $V_b = 0.1$ V, for the (a) Ag-$L_{ch}$ =2.4-nm MoS$_2$-Ag top-contact, (b) Ag-$L_{ch}$ =10.2-nm MoS$_2$-Ag top-contact, (c) Pd-$L_{ch}$ =3.1-nm MoS$_2$-Pd edge-contact, and (d) Pd-$L_{ch}$ =10.3-nm MoS$_2$-Pd edge-contact junctions (first left-top panels), we show the LDOS (first left-bottom panels), transmission $T(E; V_b = 0.1$ V$)$ (second left panels), differences in the left-electrode and right-electrode Fermi-Dirac distribution functions, $(f_L - f_R)(E; V_b = 0.1$ V$)$ (third left panels), and zoomed-in energy-resolved current spectra $T \times (f_L - f_R)(E; V_b = 0.1$ V$)$ (right panels). In all panels, the red and blue solid lines indicate $\mu_L$ and $\mu_R$, respectively. In the LDOS plots, the black dashed lines indicate the bent intrinsic CBM (upper) and VBM (lower) levels of MoS$_2$. In $T \times (f_L - f_R)$, the purple, orange, and magenta filled regions indicate the DT, FN tunneling, and TE components, respectively.